\documentclass[floatfix, reprint,
amsmath,amssymb,aps,pre,twocolumn,notitlepage
]{revtex4-1}
\usepackage{blindtext}
\usepackage{comment}
\usepackage{subfigure}
\usepackage{dcolumn}
\usepackage{amssymb}
\usepackage{float}
\usepackage{graphicx}
\usepackage{subfigure,bm}
\usepackage{color}
\usepackage{dcolumn}
\newcolumntype{x}[1]{>{\centering\arraybackslash\hspace{0pt}}p{#1}}

\begin{document}
\title{Nonequilibrium master kinetic equation modelling of colloidal gelation}
\author{Joep Rouwhorst}
\affiliation{Institute of Physics, University of Amsterdam, Science Park 904,
1098 XH Amsterdam, The Netherlands.}

\author{Christopher Ness}
\affiliation{Department of Chemical Engineering and Biotechnology, University
of Cambridge, Cambridge CB3 0AS, United Kingdom.}

\author{Theo Blijdenstein}
\affiliation{Unilever R$\&$D Vlaardingen, Olivier van Noortlaan 120, 3133 AT Vlaardingen, the Netherlands.}

\author{Alessio Zaccone}
\affiliation{Department of Physics ``A. Pontremoli'', University of Milan, 20133 Milan, Italy}
\affiliation{Department of Chemical Engineering and Biotechnology, University
of Cambridge, Cambridge CB3 0AS, United Kingdom}
\affiliation{Cavendish Laboratory, University of Cambridge, Cambridge CB3 0HE,
United Kingdom.}

\author{Peter Schall}
\affiliation{Institute of Physics, University of Amsterdam, Science Park 904,
1098 XH Amsterdam, The Netherlands.}


\begin{abstract}
We present a detailed study of the kinetic cluster growth process during gelation of weakly attractive colloidal particles by means of experiments on critical Casimir attractive colloidal systems, simulations and analytical theory. In the experiments and simulations, we follow the mean coordination number of the particles during the growth of clusters to identify an attractive-strength independent cluster evolution as a function of mean coordination number. We relate this cluster evolution to the kinetic attachment and detachment rates of particles and particle clusters. We find that single-particle detachment dominates in the relevant weak attractive-strength regime, while association rates are almost independent of the cluster size. Using the limit of single-particle dissociation and size-independent association rates, we solve the master kinetic equation of cluster growth analytically to predict power-law cluster mass distributions with exponents $-3/2$ and $-5/2$ before and after gelation, respectively, which are consistent with the experimental and simulation data. These results suggest that the observed critical Casimir-induced gelation is a second-order nonequilibrium phase transition (with broken detailed balance). Consistent with this scenario, the size of the largest cluster is observed to diverge with power-law exponent according to three-dimensional percolation upon approaching the critical mean coordination number.

\end{abstract}

\maketitle\date{\today}
\maketitle

\section{Introduction}
Under sufficient attraction, suspensions of colloidal particles undergo a
transition from a sol of individual particles to a gel state, where particles aggregate across the system leading to system-spanning rigidity~\cite{Trappe01,Lu08,Zaccarelli08,Bergenholtz03,Puertas03,Tuinier99,Blijdenstein04}. This
transition, imparting solid-like properties to colloidal suspensions at low
particle volume fraction, plays an important role in applications such as
consumer products, food technology and the processing of polymers. It also
plays an important role in the fundamental understanding of dynamical arrest in
systems of low particle concentrations, in which a rigid structure arises due to particle attractions larger than several $k_BT$ that drive the system out of equilibrium~\cite{Trappe01,Giglio}. This process has been well studied in the limit of strong attraction and vanishingly low particle volume fraction~\cite{Weitz84,Witten83,Meakin84}, where particles stick as soon as they collide, leading to diffusion-limited particle aggregation with a robust fractal dimension. In the regime of weak attractions where particles continually detach and restructure, the situation is less clear.

In this regime, system-spanning arrested structures can still form if the particle attractions are larger than some volume-fraction dependent threshold~\cite{Trappe01,Lu08}. Colloidal systems with tunable interactions, in particular colloidal depletion systems, have been used to obtain basic insight into this gelation process, and to map the phase boundaries of this transition as a function of volume fraction and particle interaction strength~\cite{Trappe01,Bergenholtz03,Puertas03,Blijdenstein04,Tuinier99,Zaccarelli05,Lu08,Eberle11}. The transition has been studied from various points of view, including dynamic arrest~\cite{Lu08,Zaccarelli08}, phase separation~\cite{Lekkerkerker02}, cluster-glass~\cite{Zaccone09} and spinodal decomposition~\cite{Lu08,Zaccarelli08,Schurtenberger1,Schurtenberger2}. As a function of the particle attractive strength, structures have been observed to change topology, becoming more compact for lower attraction due to particle rearrangement. Such particle rearrangement has been recently studied by direct particle tracking experiments on colloidal gels~\cite{vanDorn17}. As a consequence of the rearrangement, the structures become more compact and exhibit increasing fractal dimension~\cite{Shelke13,Veen12}. The resulting morphologies have important applications in product design such as cosmetics and foods, where the structure and associated rheological properties determine product stability and consumer perception. Indeed for the design of such products, it is important to understand and ultimately control the relation between attractive strength and resulting gel structure.

However, neither the onset of a space-spanning structure upon a certain attractive strength nor the attractive-strength dependent final structures are fully understood. This is the case both for physical gels, where the constituent units are colloidal particles, and chemical gels, where the constituent particles are molecules. Because interaction energies are larger than several $k_BT$, these structures form in an out-of-equilibrium process: detailed balance is typically broken as detachment and attachment rates of particles and particle clusters do no longer balance. A full understanding of the cluster evolution and resulting structure thus need to take into account the full kinetic growth process of attachment and break-up of particles and particle clusters. As this is an excessive task that is no longer analytically manageable, physically meaningful choices need to be made for simplifying approximations that in turn can be checked against experimental and simulation results. Experimentally, colloidal gels have been mostly investigated using colloidal depletion systems that allow for good realization of a tunable colloidal attraction, yet typically involve phase separation of the colloids and depletant into colloid-rich, depletant-poor, and colloid-poor, depletant-rich phases. Another complication arises due to the unavoidable effect of gravity that, despite the use of closely density-matched model systems, is known to eventually lead to sedimentation of large clusters altering their growth process. It is therefore important to compare both experiments and simulations with master kinetic equation results to identify the underlying generic process and distinguish generic from gravity-related effects.

Here we combine experiments with simulations and kinetic master equation modeling to elucidate the out-of-equilibrium process in the gelation of short-range, weakly attractive particles. We directly follow the gelation process for colloidal particles interacting with temperature-dependent critical Casimir forces that provide an effective particle attraction set by the solvent correlation length. These experiments are compared with simulations based on Langevin dynamics of spheres with short-ranged interactions that allow detailed insight into the underlying kinetic processes of cluster attachment and detachment as a function of cluster size and attractive potential. The master kinetic equation of cluster growth is then solved analytically for the case of single particle detachment, and
constant, cluster-size independent attachment
Under these approximations, the model predicts cluster mass distributions with power laws of $-3/2$ and $-5/2$, respectively, before and after gelation, which are confirmed in both experiments and simulations over the investigated range of (weak) attractive strength. These power-law distributions suggest that the observed gelation is the result of a continuous non-equilibrium phase transition. Indeed, we find both in experiments and simulations that the cluster size and correlation length diverge as a function of a single governing parameter, the mean coordination number, with an exponent according to three-dimensional percolation.

\section{Experimental method}

\subsection{Critical Casimir colloidal system}
We study gelation in systems of colloidal particles interacting with critical
Casimir forces. The attractive critical Casimir forces between the particles arise in a binary solvent close to its critical point from the confinement of solvent fluctuations between the particle surfaces~\cite{Hertlein08,Gambassi09}. The strength of the interaction is controlled by the solvent correlation length, which adjusts with temperature in a reversible and universal way~\cite{Gambassi09,Bonn09,Mohry14,Stuij17}. Thus, temperature provides a unique control parameter to tune the strength and range of the attraction, and has been used previously to induce equilibrium gas-liquid-solid phase transitions in the colloidal system~\cite{Nguyen13,Dang13}, as well as quench the system into well-defined out-of-equilibrium states~\cite{Veen12,Shelke13}. We here focus on the out-of-equilibrium case achieved for higher critical Casimir attraction, i.e. smaller $\Delta T$, for which equilibrium phases are not observed, and the system arrests in a gel. The colloids are fluorescently labeled copolymer particles made of 2,2,2-trifluoroethyl methacrylate ~\cite{Kodger15} with radius $r_0 = 1 \mu m$ and a polydispersity of 5$\%$. The particles are suspended at a volume fraction $\phi \sim 0.12$ in a binary mixture of lutidine and water, with weight fraction of lutidine $c_L = 0.25$. Sugar was added to match the solvent refractive index and density with that of the particles, while only slightly affecting the binary solvent phase diagram. We also added salt (5 mM KCl) to screen the electrostatic repulsion of the charge-stabilized particles, as in previous studies~\cite{Veen12,Stuij17}. The Brownian diffusion time based on the estimated viscosity of $\eta = 10 mPa s$ is $t_B = 7.5s$, in which the particle diffuses its own radius. Phase separation of this solvent occurs at $T_c = 31.0^\circ$C, with a critical composition of $c_c = 0.26$ as determined by systematic investigation of the solvent phase diagram over a range of compositions. To study the out-of-equilibrium gelation process, we heat the suspension to $\Delta T = T_c - T \le 1.2^\circ$C below $T_c$. We jump from room temperature to $\Delta T = 1.2, 1, 0.7$ and $0.5^\circ$C, increasing the attractive strength with each new experiment, and follow the subsequent aggregation process at the particle scale by confocal microscopy.

\subsection{Microscopic observation of gelation}

\begin{figure}
\includegraphics[width=1\columnwidth]{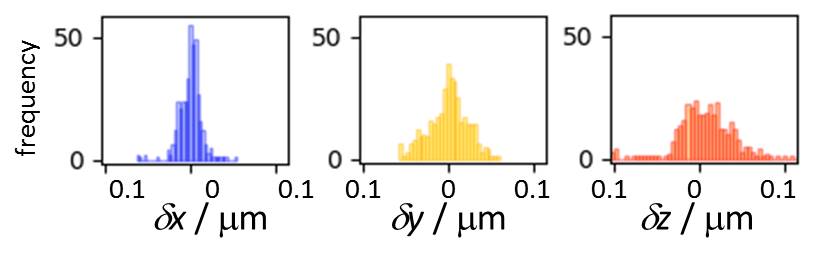}
\caption{Particle locating accuracy. Histograms of deviations of particle positions from the determined average position, in the x-, y- and z-direction.}
\label{Fig_accuracy}
\end{figure}

We use a fast laser scanning confocal microscope (Zeiss LSM 5 Live, line scanning system) equipped with a 63x lens with a numerical aperture of 1.4 to image individual colloidal particles in a $108 \mu$m by $108 \mu$m by $60 \mu$m volume. Three-dimensional image stacks with a distance of $0.2 \mu m$ between images are acquired every 60 seconds over a time interval of at least 60 minutes to follow the gelation process in three dimensions from the initial cluster formation to gelation and beyond. During this process, the temperature is kept strictly constant by using a specially designed water heating setup that controls the temperature of both the sample and the coupled oil-immersion objective with a stability of $\sim 0.01^\circ$C.
Particle positions are determined from the three-dimensional image stacks with an iterative tracking algorithm to optimize feature finding and particle locating accuracy~\cite{trackpy}. The resulting particle positions have an accuracy of $\sim$ 20nm in the horizontal and $\sim$ 40nm in the vertical direction. To show this, we used several layers of particles stuck to a cover slip, which we imaged and located repeatedly to determine histograms of particle positions, see Fig.~\ref{Fig_accuracy}. From this, we determine the positional variances $\sigma_x =$ 15nm, $\sigma_y =$ 20nm and $\sigma_z =$ 40nm. From the determined particle positions, bonded particles are identified as those separated by less than $d_0 = 2.6r$, corresponding to the first minimum of the pair correlation function. We subsequently group bonded particles into connected clusters using a clustering algorithm based on a threshold distance of $d_c = 3.5r$.\\

\section{Simulation method}

\begin{figure}
\includegraphics[width=0.475\textwidth]{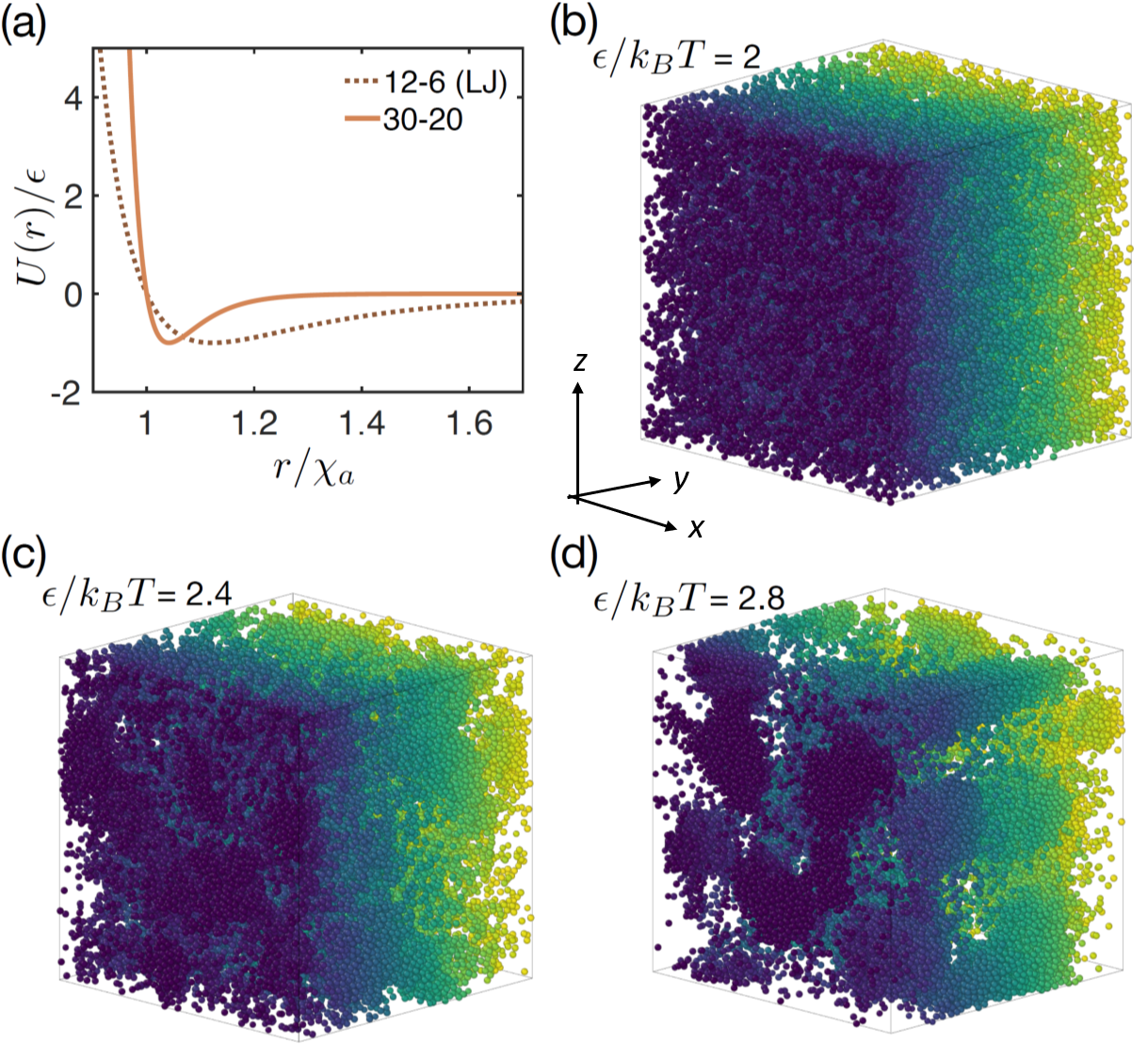}
\caption{Numerical simulations of colloidal gelation.
(a) Comparison between the 30-20 Mie potential and the square-well potential used in this work with a conventional Lennard Jones (12-6) potential. (b)-(d) Simulation snapshots near the steady state at $\epsilon/k_BT = 2$, $2.4$ and $2.8$. Color gradient from blue to yellow indicates $y$-coordinate of each particle from 0 to system size.}
\label{figure_sim1}
\end{figure}

Our simulation computes the dynamics of $N = \mathcal{O}(10^4)$ spherical
particles according to the Langevin equation
\begin{equation}
m\frac{d{v}}{dt} = -\frac{m}{\zeta}{v} - \frac{d{U}}{dr} + {f}_B(t) \text{,}
\end{equation}
for particles of mass $m$, velocity ${v}$ and coefficient of friction
$1/\zeta$, interacting with a background fluid that is implicitly modelled by
random Brownian forces ${f}_B(t)$ satisfying $\langle f_B(t)f_B(t')\rangle =
2mk_BT\delta(t-t')/\zeta$.
Particles with centres separated by a distance $r$ interact through a truncated
and shifted Mie potential with the general form
\begin{small}
\begin{equation}
U(r) = C\epsilon \left[\left(\frac{\chi}{r}\right)^{\gamma_\alpha} -
\left(\frac{\chi}{r}\right)^{\gamma_\beta} -
\left(\left(\frac{\chi}{r_c}\right)^{{\gamma_\alpha}} -
\left(\frac{\chi}{r_c}\right)^{\gamma_\beta}\right) \right] \text{,}
\end{equation}
\end{small}
where $C$ is given by
\begin{equation}
C = \left(\frac{\gamma_\alpha}{\gamma_\alpha-\gamma_\beta} \right)
\left(\frac{\gamma_\alpha}{\gamma_\beta}
\right)^{\left(\frac{\gamma_\beta}{\gamma_\alpha-\gamma_\beta} \right)}
\text{,}
\end{equation}
$\epsilon$ is a prefactor that sets the energy scale, and $r_c$ is the spatial
cut-off beyond which the interaction is not computed.
To give an attractive range comparable to that arising in the experiment due to
Casimir forces, we set $\gamma_\alpha = 30$ and $\gamma_\beta=20$. This gives
an attractive range of $(31/14)^{0.1}\chi_a\approx0.08\chi_a$, that is the
distance from the zero-crossing of the potential to its inflection point. The
potential effectively acts as a steepened Lennard-Jones potential, see
~Fig~\ref{figure_sim1}a.

We consider an equal mixture by number of particles with size ratio
$\chi_a/\chi_b = 1:1.1$, to simultaneously approximate the polydispersity of
the experimental system and to avoid
crystallisation~\cite{zaccarelli2009crystallization}. As such, three potential
neighbour permutations arise, for which $\chi$ takes values of $\chi_a$,
$\chi_b$ and $\frac{1}{2}(\chi_a + \chi_b)$. The domain has total volume $V$
(in 3D) such that the volume fraction of particles $\phi$ approximates that in
the experimental system, that is
\begin{equation}
\begin{split}
\phi &= \frac{N}{2} \frac{\pi}{6} \left(\chi_a^3 + \chi_b^3 \right)/V \\
&=0.12 \text{.}
\end{split}
\end{equation}
To check the generality of the results, we also simulate volume fractions $\phi = 6$ and $16\%$. Periodic boundaries are used in $x$, $y$ and $z$.

We operate with Lennard-Jones units throughout, setting $k_BT$ as the energy
scale, $\chi_a$ as the length scale, and letting $m =
\frac{4}{3}\pi(\frac{\chi_a}{2})^3\rho$ with $\rho$ the density scale. The time
unit is thus $t_s=\sqrt{m\chi_a^2/\epsilon}$, and we set $\zeta=t_s$ and use
$dt = 0.0025t_s$ as the numerical timestep. The simulation is implemented in
\texttt{LAMMPS}~\cite{plimpton1995fast}. Similar to the experiments, we vary
the attractive strength, $\epsilon/k_BT$, where $\epsilon$ is the prefactor of
the potential and $k_BT$ is the thermal energy. We find that this has a
dominant influence on the state of the system near steady state, as shown in
Fig~\ref{figure_sim1}(b)-(d), where we present snapshots of the system at three
values of $\epsilon/k_BT$.

To test the generality of the results predicted computationally, we additionally performed simulations on particles with an approximated square-well potential. We adopt the `continuous square-well' model described by Ref~\cite{Zeron18}, writing the potential as
\begin{equation}
U_\text{csw}(r) = \frac{1}{2}~\epsilon \left(\left(\frac{1}{r} \right)^n + \frac{1 - e^{-m(r-1)(r-w)}}{1 + e^{-m(r-1)(r-w)}} -1 \right) \text{,}
\end{equation}
using a binary form for the width of the well $w$ (potential range) to match our $\chi_a/\chi_b$ ratio as described for the Mie potential. The dimensionless well steepnesses $m$ and $n$ are set as 7000 and 700 respectively, leading to a 2nd virial coefficient (defined following Ref~\cite{Vliegenthart00}) that matches that of the Mie potential at $\epsilon/k_BT = 3$.

Systems are first equilibrated in the liquid state by setting $\epsilon/k_BT=1$
and allowing the particle trajectories to evolve for $\mathcal{O}(10^5)$ time
units. We then switch $\epsilon/k_BT$ to larger values, in practice keeping
$k_BT=1$ and varying $\epsilon$. The rate of the time evolution following this
change is set by $\zeta$. We find that percolation is absent at long subsequent
times when $\epsilon/k_BT \lesssim 2.5$.\\

\begin{figure}[b]
\includegraphics[trim={0mm 33mm 0mm 345mm},clip,width=0.475\textwidth]{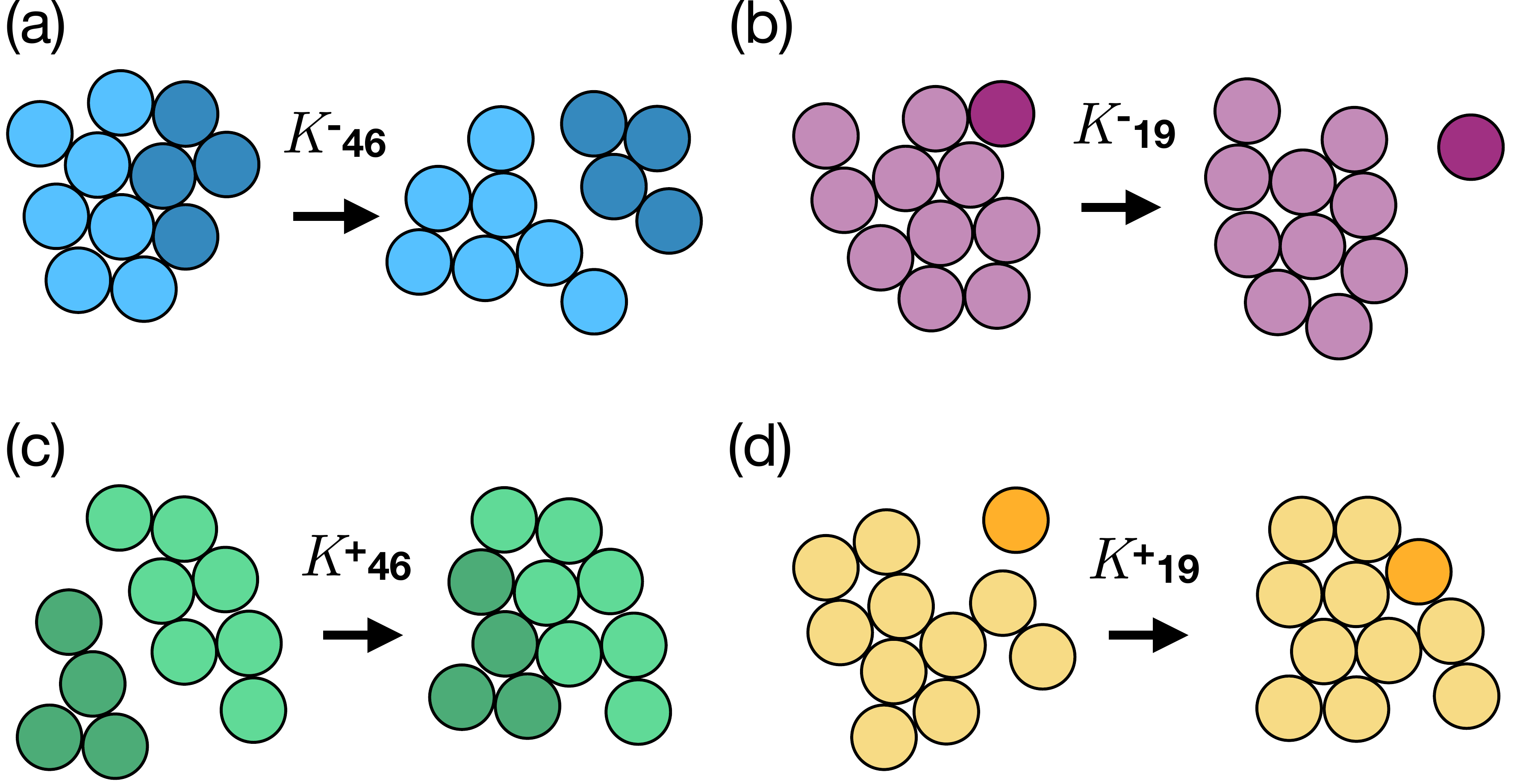}
\caption{Rate constants for dissociation and association.
Sketches in (a,b) illustrate examples of dissociation events associated with rate constant $K^-_{ij}$ for $i=4$, $j=6$ (a) and $i=1$, $j=9$ (b). In our kinetic model, $K^-_{ij}$ is set to zero if $i$, $j\neq1$. Panels (c,d) show examples of association events $K^+_{ij}$ for $i=4$, $j=6$ (c) and $i=1$, $j=9$ (d). In the model, $K^+_{ij}$ is set to be independent of $i$ and $j$. Simulation data support these assumptions for $K^-_{ij}$ and $K^+_{ij}$.
}
\label{figure_sim2}
\end{figure}

\subsection{Calculation of rate constants}

To calculate association and dissociation rates, we first define directly contacting particles as those whose centres lie within the inflection point of the potential (where $\frac{\partial^2U}{\partial r^2} = 0$), which for the values of $\gamma_\alpha$ and $\gamma_\beta$ in this case occurs at $\left(31/14\right)^{0.1}\chi_a$. Based on these criteria, we define a particle as belonging to a cluster if there exists a continuous series of direct contacts between that particle and all other particles in the cluster.
Outputting the particle coordinates with very fine time resolution then allows
us to monitor the temporal evolution of cluster sizes throughout the system as
successive dissociation and association events occur, and thus to compute the
rate constants $K_{ij}^{+/-}$ in the kinetic master equation, see below. Here,
$K_{ij}^{+}$ means the association rate of clusters that have, respectively,
$i$ and $j$ particles, while $K_{ij}^{-}$ indicates the split-up or
dissociation rate of a larger cluster into clusters of $i$ and $j$ particles. Some
examples of such dissociation and association events are illustrated in
Fig~\ref{figure_sim2}.
We determine the rate of dissociation events involving clusters of size 4, for
example, by averaging dissociation rate $K_{4j}^{-}$ over $j$. As a result, we
find that the rate of dissociation events depicted in Fig~\ref{figure_sim2}(a)
is considerably smaller than that of events as depicted in (b), where a single particle detaches from a cluster, while the rates of association events depicted in (c) and (d) are comparable.

\section{Cluster kinetic model}
The starting point of our theoretical model is a master kinetic equation for
the time-evolution of the cluster population $c_{k}$ which denotes the number of clusters with $k$ particles per unit volume, starting with a sol of isolated colloidal particles at $t=0$:
\begin{widetext}
\begin{equation}
\frac{d c_{k}}{dt}=\frac{1}{2}\sum_{i+j=k}K_{ij}^{+}c_{i}c_{j} -
c_{k}\sum_{j\geq1}K_{kj}^{+}c_{j}+\sum_{j\geq1}K_{kj}^{-}c_{j+k}-c_{k}\sum_{i+j=k}K_{ij}^{-}.
\label{eq:kinetic}
\end{equation}
\end{widetext}

In this master equation, the first term on the right hand side represents the creation of clusters with $k$ units due to aggregation of one cluster with $i$ units with another with $j$ units (where $i+j=k$); the second term represents
the``annihilation" of clusters with $k$ units due to aggregation of a cluster
with $k$ units with a cluster of any other size in the system; the third term
represents ``creation" of a cluster with $k$ units due to the breakage of a
larger cluster which splits into a cluster with $k$ units and another of $j$
units, where $j$ can take any value; the fourth term represents ``annihilation"
of a cluster with $k$ units due to fragmentation into two fragments $i$ and
$j$, subjected to mass balance. There is a set of $k$ such differential equations for each cluster size, thus forming a system of ordinary differential equations that has to be solved in order to obtain the cluster mass distribution as a function of time. The rate coefficients $K_{ij}^{+}$ represent the aggregation rates between two clusters $i$ and $j$, whereas terms of the type $K_{ij}^{-}$ represent the fragmentation rates of a cluster $i+j$ into two fragments $i$ and $j$. Here fragmentation is due solely to
thermally-activated breaking of bonds across the aggregate.

The above master equation in its most general form can only be solved
numerically. However, analytical solutions are possible for certain models.
Based on physical intuition, in colloidal aggregation particles that are on the
surface of the cluster can more easily detach by thermal motion, since they are
bonded to a smaller number of other particles beneath, whereas particles in the
inner part of the cluster have many more connections and therefore those bonds
are much more difficult to break by thermal energy.

This consideration motivates us to consider the following schematic model of
aggregation relying on two basic assumptions: i) aggregation takes place
between any two clusters of arbitrary size, with a rate constant $K_{ij}^{+}$
independent of cluster size; ii) dissociation involves detaching of dangling
particles only (1-fold coordinated). In other words, breakup events leading to
two fragments, each of them larger than one particle, are excluded.
Under these assumptions, the rate coefficients are given by:
\begin{equation}
\label{eq:rateconstants}
\begin{split}
K_{ij}^{+}&=const,~~~\forall~i,j\\
K_{ij}^{-}&=\lambda K_{ij}^{+},~\mathrm{if}~i=1,~\mathrm{or}~j=1\\
K_{ij}^{-}&=0,~\mathrm{if}~i\neq1,~\mathrm{or}~j\neq1 .
\end{split}
\end{equation}

Clearly, the last condition breaks the detailed balance: there is no linear
dependence between aggregation and fragmentation rates for all processes
involving $i$ and $j$ both larger than unity, or in other words these
aggregation processes are irreversible. The basic implication of this condition
is that any stationary state (for which cluster mass distribution reaches a
steady-state in time) is a nonequilibrium stationary state. In turn, a
transition from one such nonequilibrium steady-state to another nonequilibrium
state is a nonequilibrium phase transition.

Upon using Eq.~\ref{eq:rateconstants} in Eq.~\ref{eq:kinetic} and introducing
the generating function (a procedure similar to a discrete Laplace
transformation) $C(z,t)=\sum_{j\geq1}(z^{j}-1)c_{j}(t)$, where $z$ is a dummy
variable as usually defined in generating functions, the system of ordinary differential equations is reduced to the following Riccati equation:
\begin{equation}
\frac{dC}{dt}=C^{2}+2\lambda\frac{1-z}{z}C+2\lambda\frac{(1-z)^{2}}{z} N(t)
\end{equation}
where $N(t)=\sum_{j\geq1}c_{j}(t)$ and we took $K_{ij}^{+}=2$ for ease of
notation and without any loss of generality~\cite{majumdar1,majumdar2}.
At steady-state, $dC/dt=0$ or $t\rightarrow\infty$, the second-order algebraic
equation is solvable, and differentiating $C$ with respect to $z$ and setting
$z=1$ gives $N$ as a function of $\lambda$. A continuous phase transition at
the critical point $\lambda_{c}=1$ is found, which separates the sol state with
$N=1-(2\lambda)^{-1}$ from the gel (spanning network) state with $N=\lambda/2$.

\begin{figure*}
\includegraphics[width=0.8\textwidth]{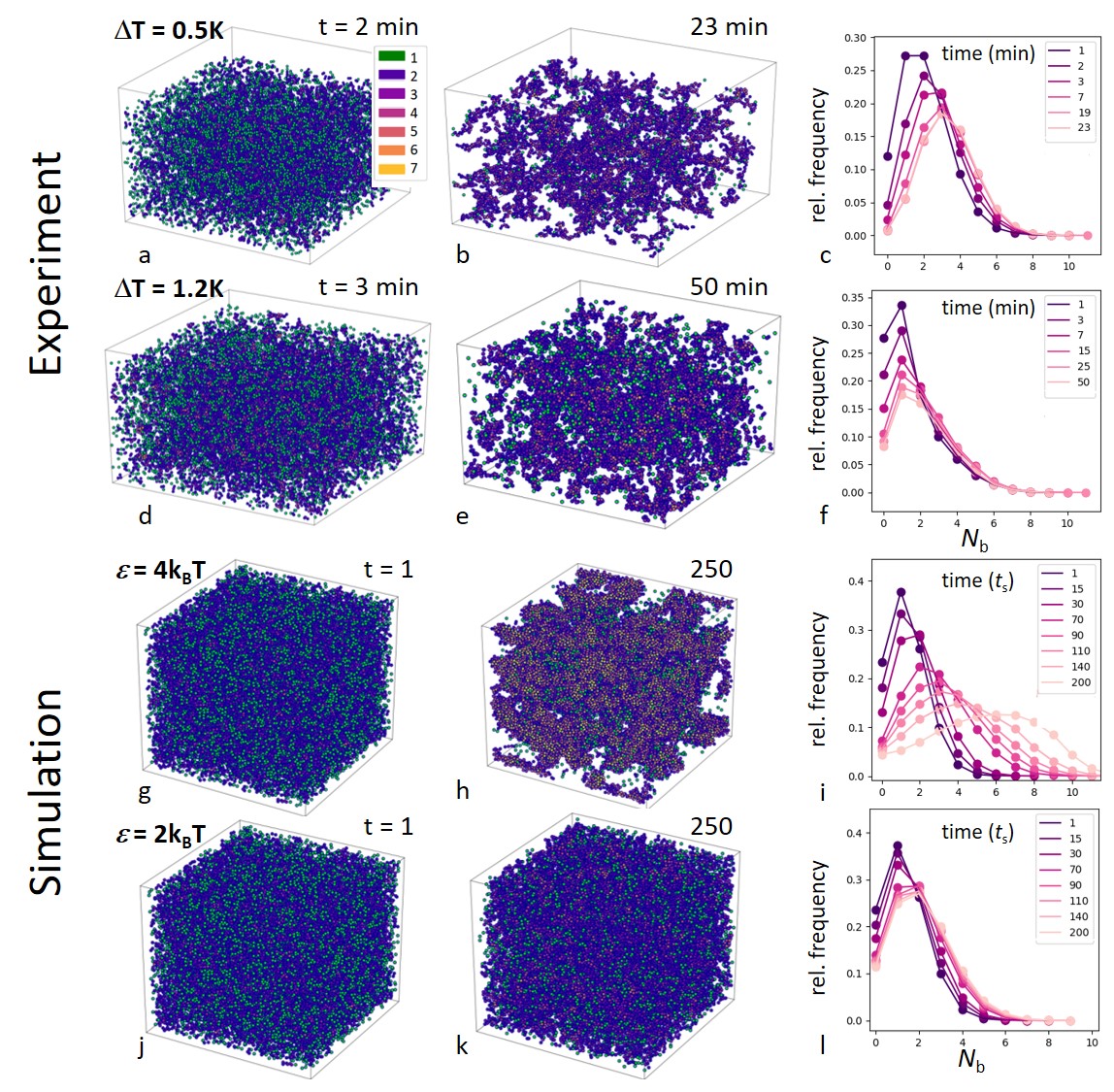}
\caption{Observation of gelation in experiments and simulations. (a)-(f) Reconstructions and bond histograms of aggregating colloidal particles in experiments at $\Delta T = 0.5K$ (a-c) and $\Delta T = 1.2K$ (d-f). Particle color indicates the number of bonded neighbours, see color bar. (c,f). Bond histograms for $\Delta T = 0.5K$ (c) and $1.2K$ (f). Color indicates aggregation time in minutes (see legend). With time, distributions shift towards higher number of bonds, indicating increasingly connected structures. The aggregation time interval of $t=$~50min corresponds to 400$t_B$. (g-l) Snapshots and bond histograms of colloidal aggregation in simulations at $\epsilon = 4 k_BT$ (g-i) and $2 k_BT$ (j-l). Time is given in units of $t_s$, indicated in color in (i) and (l), see legend. Similar trends as in the experiments are observed in the aggregate topology and bond distributions.}
\label{Fig1}
\end{figure*}

This is best seen by looking at the cluster mass distribution (CMD). By
expanding $C(z)$ in powers of $z$ one obtains the cluster mass distribution in
the sol phase and in the gel phase. In the pre-critical sol phase, the
power-law is accompanied by an exponential cut-off~\cite{majumdar1,majumdar2},
\begin{equation}
c_{k}\left(t\rightarrow\infty\right)\sim k^{-3/2}e^{-k/k_{c}}.
\end{equation}
The presence of the exponential cut-off implies that all clusters are finite in
size. However, the cut-off size $k_{c}$ diverges at $\lambda\rightarrow1^{+}$,
according to ~\cite{majumdar1,majumdar2}
\begin{equation}
\label{eq5}
k_{c}=\left\{2\log\left(\lambda/\lambda_{c}\right)-\log\left[2\left(\lambda/\lambda_{c}\right)-1\right]\right\}^{-1}.
\end{equation}

In the gel phase $\lambda \geq 1$, the steady-state cluster mass distribution is
\begin{equation}
c_{k}\left(t\rightarrow\infty\right)\sim k^{-5/2},
\end{equation}
now without an exponential tail, which signals the existence of a giant
system-spanning cluster via the divergence of the first-moment of the
distribution.

Hence, this model predicts gelation as a continuous (second-order)
phase transition, with a cluster-mass distribution that exhibits two distinct
power-law exponents, namely $\tau=-3/2$, with an exponential tail, in the sol
phase, and $\tau=-5/2$, without the exponential tail, in the gel phase.

Furthermore, we can link the breakage rate with the attraction energy.
For a purely attractive potential well,the breakup rate $K_{ij}^{-}$ is given
by the Kramers escape rate of the individual particle detaching from the
cluster. The precise shape of the potential in the experimental system is not known, but we can still get an order of magnitude estimate for a simple square-well attraction,
$K_{ij}^{-}=\left(D/\delta^{2}\right)e^{-\epsilon/k_{B}T}$, where $D$ is the
diffusion coefficient of one particle, $\delta$ is the range of attraction, and
$\epsilon$ is the interaction potential~\cite{Zaccone_jor}. Together with the
condition $\lambda_{c}=1$, in units of $c_{0}$, and with the assumption that
the aggregation rate is diffusion-limited,
$K_{ij}^{+}=\left(8/3\right)k_{B}T/\mu$, with $\mu$ the solvent viscosity, this
leads to a rough estimate for the critical attraction energy~\cite{Zaccone_jor}

\begin{equation}
\label{eq6}
-\epsilon_{c}/k_{B}T \sim \log\left[12 \left(\delta/a \right)^{2}
\phi_{0}\right]
\end{equation}
where $\phi_{0}=\left(4/3\right)\pi a^{3} c_{0}$ is the solid volume fraction.
This equation provides an order of magnitude estimate of the minimum attraction energy between two colloidal particles to have gelation in steady-state. For example, at
a volume fraction at $\phi_{0}=0.12$ and $\delta/a=0.08$, this formula gives
$\epsilon_{c}\simeq 4.7 k_{B}T$. This value is in the same order of magnitude (and within a factor 2) of the value $\epsilon_{c}\sim 2.5 k_{B}T$ determined in simulations.

Assuming that the scaling hypothesis holds, i.e. that the various quantities
are power-law functions of the distance from the critical point, as is expected
for continuous phase transitions (even though this cannot be rigorously proven
because a suitable free energy cannot be defined in this case), the
hyperscaling relation of critical phenomena~\cite{Stauffer} is then also
expected to hold: $\tau=(d/d_{f})+1$. Here, $\tau$ is the power-law exponent of
the CMD at the critical point, $d_f$ is the fractal dimension of the system
(hence of the clusters forming the gel) and the spatial dimension $d=3$. Using the critical exponent $\tau=5/2$ gives the prediction $d_{f}=2.0$ for the fractal dimension, also to be verified below in
comparison with experiments and simulations. One should note that this estimate does not account for ageing phenomena due to restructuring of the clusters into denser aggregates which typically leads $d_f$ to increase at low attractive strength~\cite{Veen12,Shelke13}.

The above model predictions for $\tau$ and $d_{f}$ can now be tested
experimentally on a well-controlled system. Furthermore, the key hypothesis in
Eq.~\ref{eq:rateconstants} that only individual particles (on the surface of
the clusters) break off due to thermal motion can be tested in numerical
simulations.

\begin{figure*}
\includegraphics[width=0.8\textwidth]{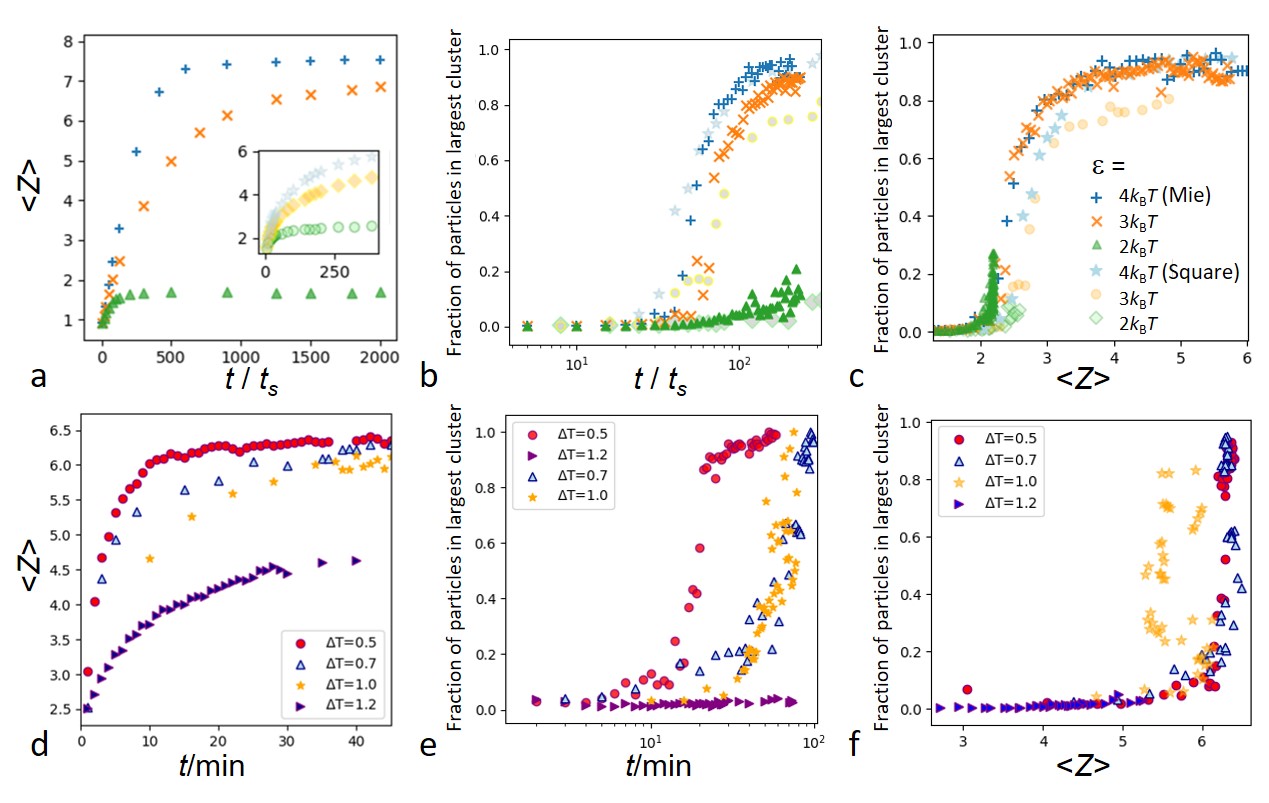}
\caption{Evolution of the cluster size and mean coordination number in simulations (a-c) and experiments (d-f). (a) Mean coordination number as a function of time in simulations (main panel: Mie potential, inset: square-well potential). (b) Fraction of particles in the largest cluster as a function of time in simulations. (full symbols: Mie potential, faint symbols: square-well potential) (c) Fraction of particles in the largest cluster as a function of mean coordination number. (d-f) Mean coordination number and fraction of particles in the largest cluster in experiments.}
\label{Fig3}
\end{figure*}

\begin{figure}
\includegraphics[width=0.7\columnwidth]{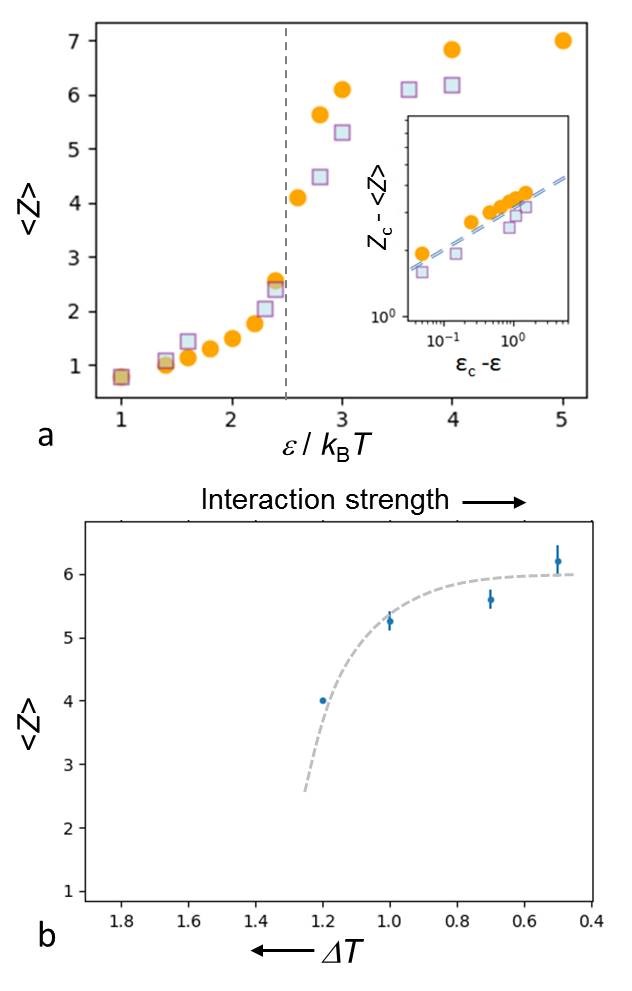}
\caption{Late-stage saturation value of the mean coordination number as a
function of attractive strength in simulations (a) and as a function of
temperature difference $\Delta T$ in experiments (b). The two data sets in (a) indicate Mie potential (yellow dots), and square-well potential simulations (gray squares). Inset shows the approach of the critical mean coordination number as a function of attractive strength difference to the critical attraction $\epsilon_{c}$, when approaching $\epsilon_{c}$ from below. Critical scaling with exponent $\sim 1/6$ is
observed.}
\label{Fig4}
\end{figure}

\section{Results}

\subsection{Cluster growth and bond evolution}
We follow the growth of particle clusters in experiments under well-defined critical Casimir attractions, which we induce by heating the suspension to well-defined temperatures $\Delta T$ below $T_c$. Reconstructions show the time evolution of the colloidal system at $\Delta T=0.5K$ in Fig.~\ref{Fig1}a and b. Particle color indicates the local coordination number, i.e. the number of bonded neighbors of a particle. Initially, particles exhibit no or only very few bonded neighbors, while at later stages, as clusters grow and the particles become increasingly connected, the number of bonds increases. To quantify this change of bonding configuration, we plot the relative frequency as a function of number of bonds $N_b$ in Fig.~\ref{Fig1}c. Initially, low-bonded configurations are most prominent, indicating prevalence of monomers and small clusters. As clusters grow, the bond probability distribution shifts to the right, where it eventually saturates. Results for $\Delta T = 1.2K$ corresponding to significantly lower attractive strength are shown in Fig.~\ref{Fig1}d-f. In this case, the particles no longer reach a space-spanning structure as shown by the absence of a network-like structure in the late-stage reconstruction in Fig.~\ref{Fig1}e: particles are more distributed over space, leading to disconnected clusters, and non-bonded or single-bonded particles. This is reflected in the probability
distributions of bonds shown in Fig.~\ref{Fig1}f that are shifted to the left
with respect to those in Fig.~\ref{Fig1}c. The data shows a trend towards
lower $N_b$, which remains low over time, indicating fewer bonds, and less well-connected particles.

Similar change of topology and bond configuration is observed in the
simulations performed at the different attractive strength. Reconstructions show the growing aggregates at $\epsilon = 4k_BT$ in Fig.~\ref{Fig1}g and h, where again the number of bonds per particle is indicated with color. Similar to experiments, the distribution of bonds shift to the right as clusters grow (Fig.~\ref{Fig1}i), reflecting an increasing fraction of bulk particles that sit deeper in the structure. At lower attractive strength, $\epsilon = 2k_BT$, space-spanning structures no longer form, as shown by the reconstructions in Fig.~\ref{Fig1}j and k, in qualitative agreement with experiments.

\subsection{Coordination number}

To investigate the emergence of space-spanning structures as a function of the growing number of bonds, we define the mean coordination number, $\langle Z \rangle = (1/N) \sum_{i=1}^{N} N_{b,i}$, where $N_{b,i}$ is the number of bonds of particle $i$, and $N$ is the number of particles. The coordination number increases monotonically as structures grow as shown in Fig.~\ref{Fig3}, rising to a unique, attractive-strength dependent value, where it saturates. This growth of the mean coordination number reflects the emergence of increasingly connected clusters spanning increasing portions of
space, as illustrated by the reconstructions in Fig.~\ref{Fig1}. The same trend is observed in the simulations based on the square-well potential, as shown in the inset of Fig.~\ref{Fig4}a. We plot the fraction of particles in the largest cluster as a function of time for the different attractive strengths in Fig.~\ref{Fig3}b. At sufficiently large attractive strength, this fraction grows sharply until the largest cluster has absorbed almost all particles. The rate of growth depends on the attractive strength: Higher attraction leads to faster, lower attraction to slower growth. This is true for both the Mie-potential as well as the square-well potential simulations, showing that this behavior is robust. At the lowest attraction, the largest cluster no longer absorbs a major fraction of particles and its size remains rather limited. Indeed, the real-space reconstructions show that in this case, clusters no longer span the field of view, in contrast to the situation at larger attraction.

Remarkably, when we plot the size of the largest cluster as a function of mean coordination number, we find that all Mie potential curves collapse onto a single master curve, indicating a common underlying mechanism of growth, characterized by a single parameter, the mean coordination number, see Fig.~\ref{Fig3}c. We hence consider the mean coordination number as the order parameter of the growth and gelation transition process, which is justified in view of the linear relation between $\langle Z \rangle$ and the shear modulus $G$~\cite{Zaccone11}, with the latter being identically zero in the sol phase and non-zero (positive) in the gel phase. The square-well potential data shows the very same trend, albeit slightly shifted, due to the slightly modified definition of nearest neighbors for this different form of the potential. We will see below that this will not change the overall scaling, which turns out to be identical. Interestingly, while the different attractive strengths follow the same master curve, data for the lowest attractive strength remain limited to the lowest part of the curve, as the mean coordination number saturates at its attractive-strength dependent value (Fig.~\ref{Fig3}a).

\begin{figure}
\includegraphics[width=0.6\columnwidth]{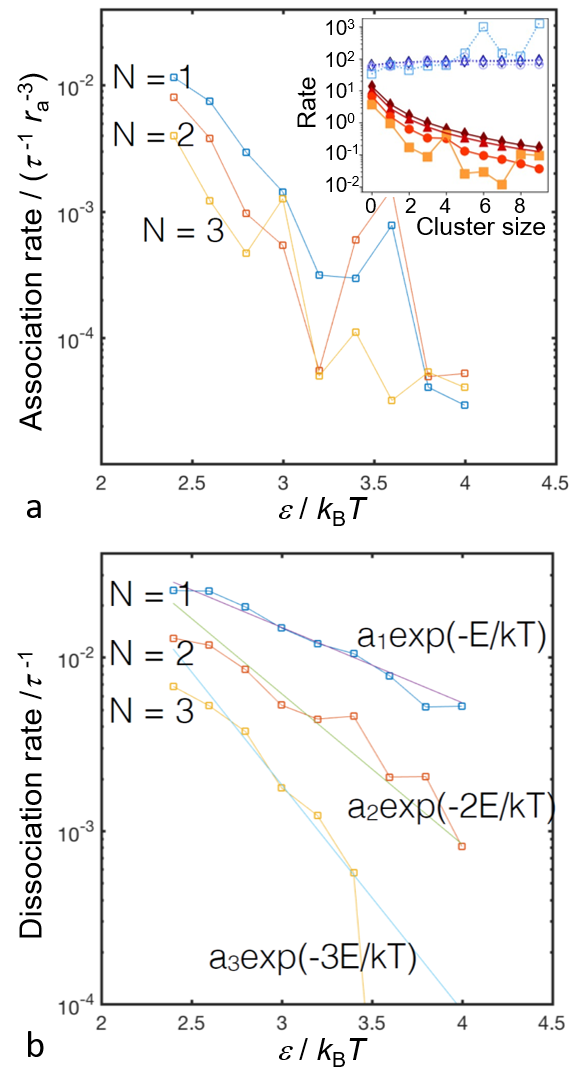}
\caption{Association (a) and dissociation rate (b) in simulations (Mie potential) as a function of attractive strength for clusters of $N=$ 1, 2, and 3 particles. The dissociation rates depend strongly on the size of the detaching cluster, decreasing with attractive strength proportional to $exp(-NE/k_BT)$, with $N$ the number of particles in the dissociating cluster. Inset in (a) shows association (blue symbols, top) and dissociation rates (red symbols, bottom) as a function of cluster size. Curves from top to bottom indicate increasing $\epsilon/k_\textrm{B}T =$ 2 (triangle), 2.2 (diamond), 2.4 (dot), and 2.6 (square), across the gelation transition.}
\label{Fig5}
\end{figure}

Similar behavior is observed in the experiments. The mean coordination number
as a function of time shows a similar attractive-strength dependent growth and saturation (Fig.~\ref{Fig3}d), while the fraction of particles in the largest cluster shows a similar rapid increase (Fig.~\ref{Fig3}e). Specifically, the largest cluster again absorbs almost all particles, given sufficient attractive strength. For the lowest attraction in experiments ($\Delta T = 1.2K$), the system does not gel yet and the largest cluster remains very small. For all higher attractions, the system gels, and the data shows divergence of the largest cluster. This is shown by plotting the size of the largest cluster as a function of mean coordination number in Fig.~\ref{Fig3}f, where the data for the two highest attractive strengths ($\Delta T =0.5$ and $0.7K$) overlap; for the next weakest attraction ($\Delta T=1.0K$), the gel is just marginally stable and is most affected by gravity, leading to some deviation at later stages (yellow stars), due to sedimentation of the largest cluster, as confirmed by direct observation of the cluster in the confocal microscope. At the weakest attraction ($\Delta T=1.2K$, violet lying triangle), the data still falls on top of the Master curve, but no longer curves up as the clusters remain small and the system does not gel. Nevertheless, the data shows good collapse (besides the discrepancies caused by gravity) also in the experiments, and this data collapse supports a common mechanism of growth, governed by the mean coordination number. 

\begin{figure*}
\includegraphics[width=0.8\textwidth]{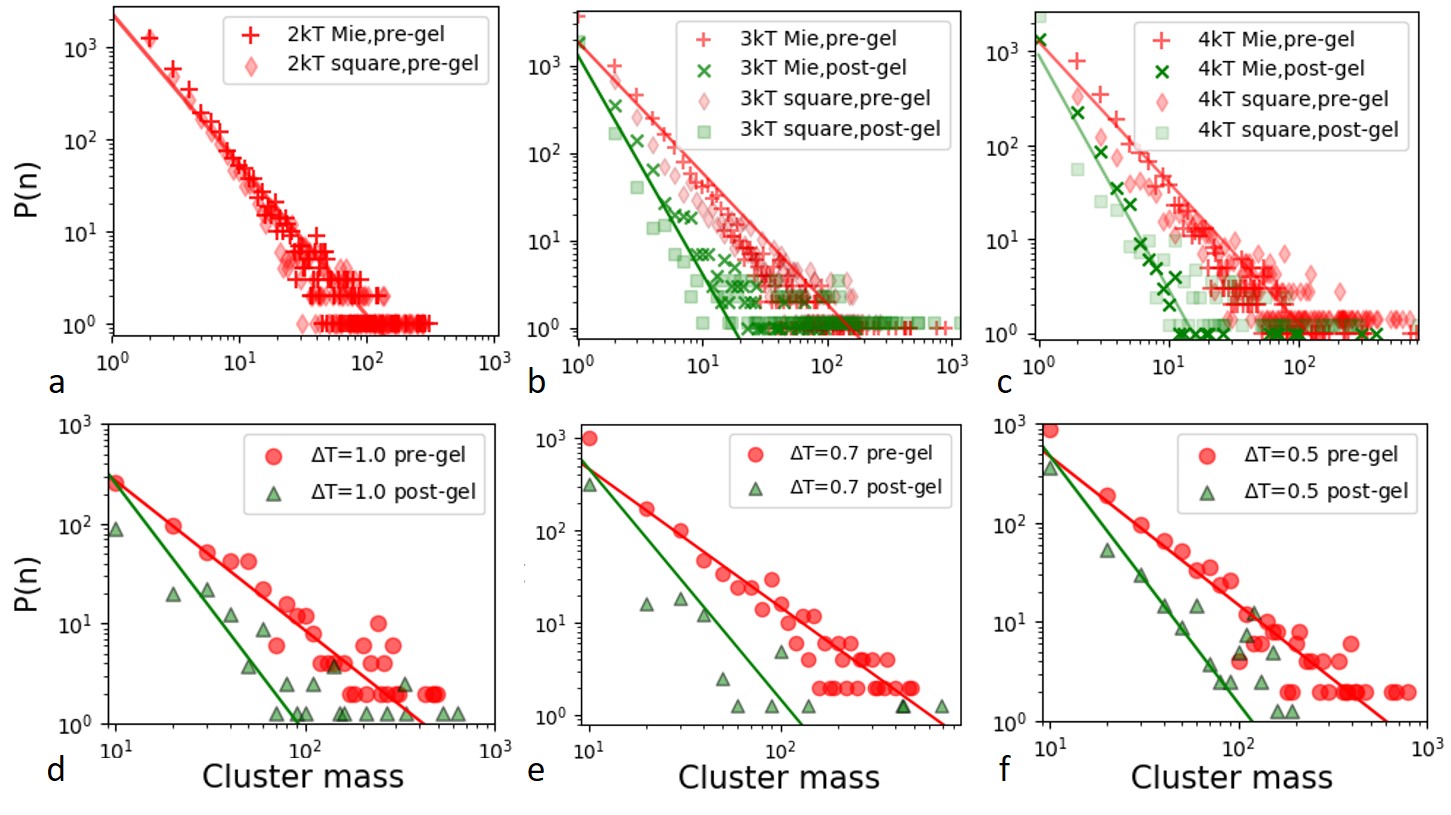}
\caption{Cluster mass distributions in simulations (a-c) and experiments (d-f), for various attractive strength, before (red "+", diamonds and dots) and after gelation (green "$\times$", squares and triangles).}
\label{Fig6}
\end{figure*}

To investigate the occurrence of gelation as a function of attraction in more detail, we plot the mean coordination number $\langle Z \rangle$ (i.e. the saturation value that $\langle Z \rangle$ approaches at long times) as a function of attractive strength in Fig.~\ref{Fig4}. The steady-state mean coordination number increases monotonically with attraction, most strongly between 2 and 4$k_BT$, after which it approaches a value above 6, slightly higher than the isostatic hard-sphere value ($2d = 6$)~\cite{Zaccone11}. We find that system-spanning clusters occur at attractive strength larger than $\epsilon_{c} \sim 2.5k_BT$ (vertical dashed line), with a critical mean coordination number of $Z_{c}\sim 3$. Both simulation data show very similar behavior (with slightly different values of $\langle Z \rangle$, due to the different potential form). Furthermore, the onset of gelation upon approaching this critical mean coordination number is identical: We plot the coordination number difference, $Z_{c} - \langle Z \rangle$ as a function of attractive strength difference upon approaching $\epsilon_{c}$ from below, $\epsilon_{c} - \epsilon$, in Fig.~\ref{Fig4}a inset. Both data sets suggest a power-law approach of the critical coordination number according to $Z_{c} - \langle Z \rangle \propto (\epsilon_{c} - \epsilon)^\alpha$, with exponent $\alpha \sim 1/6$.

Qualitatively, a similar plot of mean coordination number versus attraction is observed in experiments. However, in the experiments, the late stages are affected by gravity. We therefore chose values of $\langle Z \rangle$ for an intermediate stage at which the coordination number has reached close saturation, but gravitational disturbances are still small. The resulting mean coordination number as a function of the parameter $\Delta T$ controlling the attraction is shown in Fig.~\ref{Fig4}b. The data shows a qualitatively similar increase of $\langle Z \rangle$ as in the simulations, which is also in qualitative agreement with data on colloid-polymer mixtures presented in~\cite{vanDorn17}. A full quantitative relation would require detailed elaboration of the critical Casimir force as in ~\cite{Stuij17}, and a gravity-free environment, which is not within reach of this study.

\subsection{Association and dissociation rates}

The evolution of the mean coordination number is the result of dynamic association and dissociation processes governing the non-equilibrium growth of the structure: Their balance determines the kinetic pathway of growth of the clusters. To obtain insight into the kinetics of attachment and break-up processes, we determine association and dissociation rates in the simulations (see simulation section), and plot them as a function of attractive strength $\epsilon$ in Fig.~\ref{Fig5}. We show data for the association and dissociation of clusters consisting of 1, 2 and 3 particles. The dissociation rate depends systematically on cluster size: it decreases with $\epsilon$ roughly in an exponential manner as is expected for simple Arrhenius behavior, while the exponent reflects the number of broken bonds, leading to steeper exponential decay for larger clusters, in agreement with the experimental observations in~\cite{vanDorn17}. This is most clearly seen by the decreasing dissociation rate with cluster size in the inset (bottom, red data). At the same time, the association rate is not so much affected by the cluster size (top, blue data). For the regime of attractive strength sufficiently large for gelation to occur ($\epsilon \sim 3-5 k_BT$) the dissociation of single particles is thus significantly more likely than two particle or three-particle clusters, while the association rate depends much less on the cluster size. 
In the cluster kinetic model, to be analytically tractable, we assume the limit, in which only single-particle detachment occurs, and the association rate is independent of the cluster size.

\subsection{Cluster kinetic modeling and non-equilibrium phase transition}

We can now model the time evolution of clusters based on the full kinetic growth equation eq.~\ref{eq:kinetic} describing the joining and splitting up of clusters. In order to do this analytically, we solve the system of coupled differential equations under the simplifying assumptions (eq.~\ref{eq:rateconstants}) that the association rate is constant (independent of the cluster size), and the dissociation is governed by single-particle detachment alone, neglecting detachment of any larger clusters consisting of two or more particles. As shown in Fig.~\ref{Fig5} and discussed above, these assumptions are reasonably well fulfilled in the relevant attractive-strength regime. In this case, the model predicts power-law cluster mass distributions with exponents $-3/2$ before percolation, and $-5/2$ thereafter. To compare with simulations and experiments, we plot cluster-mass distributions before and after percolation for the different attractive strength in Fig.~\ref{Fig6}. In all cases, the data follows closely the predicted power-law slopes before and after gelation. Specifically, we find slopes of $1.60 \pm 0.12$ and $1.65 \pm 0.21$ for the Mie simulations at 3 and $4k_BT$, respectively, before gelation, and slopes of $2.30 \pm 0.35$ and $2.62 \pm 0.26$ after gelation, consistent with the predictions of the model. Similar consistency is obtained for the square-well potential simulations, and even for the lowest attractive strength, for which no gelation occurs, and cluster mass distributions can only be determined before gelation (Fig.~\ref{Fig6}a). While thus the scatter of the data does not allow to precisely pinpoint the power-law slopes, the data is consistent with the model predictions. As the model is based on idealized limits that are not precisely fulfilled in the simulations, an exact agreement between predicted and measured slopes may not be expected. Equally good consistency is observed in the experiments, see Fig.~\ref{Fig6}d-f. The data is well described by the predicted power-law slopes for all attractive strengths. For example, at $\Delta T = 0.7K$, we determine slopes of $1.6 \pm 0.2$ before gelation, and $2.4 \pm 0.6$ after, while at $\Delta T = 0.5K$, we find $1.55 \pm 0.15$ before gelation, and $2.1 \pm 0.5$ after, both consistent with the predictions, while the scatter does not allow us to make this statement more precise. On the other hand, the reasonable agreement of all attractive strength, and of both simulation models and experiments points to some underlying generality. The model thus gives reasonable predictions of cluster mass distributions over the relevant range of attractive strength, where single-particle detachment prevails. For attraction lower than the critical value, clusters do not grow to sufficient size and the mean coordination number does not increase to a high enough value for percolation to be achieved.

\begin{figure*}
\includegraphics[width=0.7\textwidth]{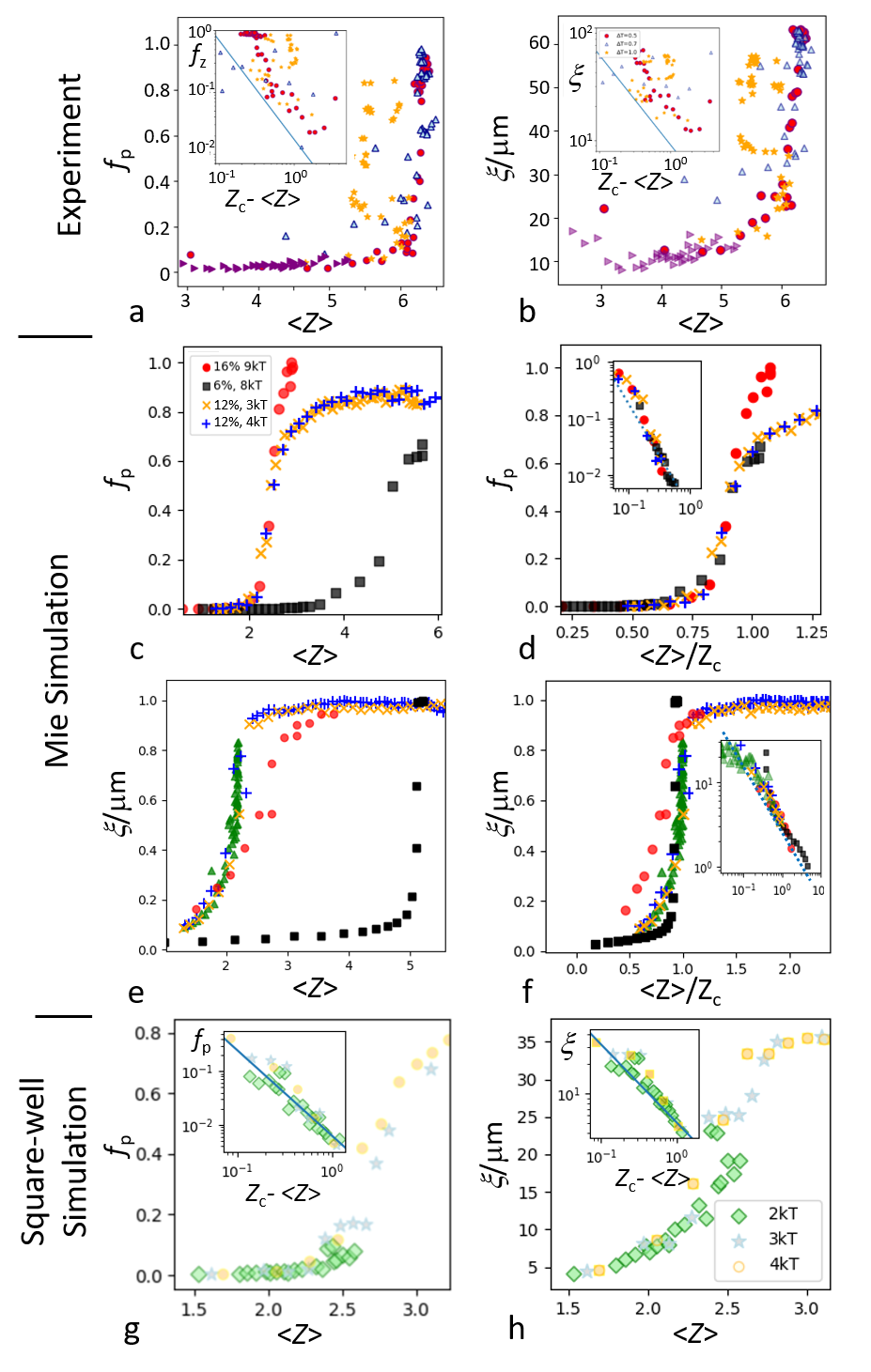}
\caption{Evolution of cluster size and correlation length in experiments (a,b), Mie-potential simulations (c-f), and square-well simulations (g,h). (a) Fraction of particles in the largest cluster, $f_p$, as a function of the total mean coordination number in experiments. Inset: Divergence of $f_p$ upon approaching the critical coordination number $Z_c$. Line indicates exponent $-1.6$ (b) Correlation length $\xi$ as a function of the mean coordination number for experiments. Inset: Divergence of $\xi$ upon approaching the critical coordination number $Z_c$. Line indicates exponent $-0.8$. (c,d) Fraction of particles in the largest cluster, $f_p$ as a function of the mean coordination number $\langle Z \rangle$ (c) and normalized mean coordination number $\langle Z \rangle / Z_c$  (d) in Mie-potential simulations for different volume fractions and attractions, see legend. Inset: Divergence of $f_p$ upon approaching the critical coordination number $Z_c$. Dashed line indicates exponent $-1.6$ (e,f) Correlation length, $\xi$, as a function of mean coordination number $\langle Z \rangle$ (e) and normalized mean coordination number $\langle Z \rangle / Z_c$ (f) for Mie simulations. Inset: Divergence of $\xi$ upon approaching the critical coordination number $Z_c$. Dashed line indicates exponent $-0.8$ (g,h) Same quantities for simulations using a square-well potential. Good agreement with the other simulations and experiments is observed.}
\label{Fig9}
\end{figure*}

A crucial prediction of the model is a nonequilibrium critical point, at which cluster sizes diverge, and the largest cluster spans the entire system. Indeed, the rapid growth of the largest cluster absorbing almost all particles (Fig.~\ref{Fig3} b,e) and the observation of power-law cluster size distributions support this scenario. To address this crucial point directly, we investigate the divergence of the largest cluster and the corresponding correlation length upon approaching the critical mean coordination number $Z_{c}$. We compute the correlation length $\xi$ of clusters using $\xi^2 = 2 \sum_i R_{gi}^2 N_i^2 /\sum_i N_i^2 $ where $R_{g,i}$ is the radius of gyration of clusters of size $N_i$ \cite{Stauffer}. We then plot the fraction of particles in the largest cluster, and the correlation length $\xi$ as a function of the order parameter $\langle Z \rangle$ in Fig.~\ref{Fig9} for all attractive strength, experiments (top), and simulations (center and bottom). Indeed, cluster sizes diverge as the mean coordination number reaches the critical value, $Z_{c}$. For the lowest attraction in experiments ($\Delta T = 1.2K$), the system does not gel yet and the largest cluster remains very small. For all higher attractions, the system gels, and the data shows divergence of the largest cluster. This is shown by plotting the cluster size and correlation length as a function of distance to the critical coordination number, $Z_{c} - \langle Z \rangle$, in the inset. We find for the highest attraction ($\Delta T = 0.5K$) a divergence with exponent $\gamma = 1.67 \pm 0.14$, while the correlation length diverges as $\nu = 0.78 \pm 0.09$. Both are consistent with predictions from three-dimensional percolation theory of $\gamma \sim 1.6$ and $\nu \sim 0.8$. 
Some scatter is observed in the experiments at low attraction ($\Delta T = 1K$, yellow stars) where the gel is just marginally stable and thus most strongly affected by gravity, and a slope cannot be determined. 
Similar divergence is observed in the simulations, which are also consistent with each other: We show the size of the largest cluster in the Mie simulations at a few different attractions and volume fractions in Fig.~\ref{Fig9}c. Divergence at a volume-fraction dependent critical coordination number $Z_c$ is observed. However, the curves collapse when we scale the coordination number by $Z_c$, as shown in Fig.~\ref{Fig9}d. The data show a characteristic scaling with slope $\gamma = 1.7 \pm 0.2$, again consistent with percolation theory. Similar scaling collapse is observed for the correlation length, see Fig.~\ref{Fig9}e and f. Again, divergence is observed at the volume-fraction dependent $Z_c$, which however collapses onto similar scaling with exponent $\nu = 0.78 \pm 0.06$, consistent with 3D percolation. This scaling is also observed in the square-well simulations, see Fig.~\ref{Fig9}g and h, where we show a where we show results for the volume fraction $\phi=12\%$, and different attractive strength.
Altogether, these data suggest that the observed short-range attractive gelation is associated with a continuous non-equilibrium phase transition. The collapse of the different attractions and volume fractions studied, and its consistent observation in experiments, Mie- and square-well simulations suggests that this finding may be more general, in line with recent simulations of the jamming of attractive spheres~\cite{Tighe2018}.

\section{Conclusions}

By investigating the growth kinetics of clusters close to gelation of short-range, weakly attractive colloidal particles in experiments, simulations and cluster kinetic modeling, we have identified a general underlying kinetic growth mechanism that is independent of the attractive strength and volume fraction in the investigated regime. While the bond and cluster evolution are clearly attractive-strength dependent, exhibiting higher coordinated, faster growth of structures for higher attractive strength, we find that all growth curves can be uniquely parameterized in terms of the mean coordination number, i.e. the mean number of bonds per particle, as a function of which all growth curves overlap. Gelation occurs when the steady-state value of this mean coordination number reaches a volume-fraction dependent critical value, above which particle clusters grow to system size. The mean coordination number, which results from a dynamic balance of association and dissociation of particles and particle clusters, reaches its critical value at gelation in a critical fashion as a function of the attractive strength. Detailed analysis of the association and dissociation rates show that in the studied weakly-attractive regime, dissociation is dominated by single-particle detachment from clusters, while association appears to be more independent of the size of the attaching cluster. Assuming the limit of only single-particle dissociation and equal association probability for all cluster sizes, we solve the general kinetic equation of cluster growth analytically. The analytical solution shows the occurrence of a critical (percolation) point, below which clusters remain finite, and above which the cluster sizes diverge. Indeed, in both experiments and simulations, we find divergence of cluster sizes and correlation lengths consistent with three-dimensional percolation theory as a function of the underlying order parameter, the mean coordination number, that characterizes the connectivity of the structure. 

This divergence is observed in all our data, colloidal experiments with different critical Casimir attractive strengths, and in Mie- and square-well potential simulations of different attractive strength, and volume fractions in the regime of $6-16\%$, indicating that it may be a general feature of the gelation of short-range, weakly attractive particles. As the range of attraction of our work is similar to the work in~\cite{Lu08,Zaccarelli08}, the question arises how the results are related. We emphasize that we investigate our system towards the non-equilibrium gel state, while Refs.~\cite{Lu08,Zaccarelli08} show cluster distributions in the equilibrium cluster phase. Hence, the description here starts from a different angle, from a purely nonequilibrium kinetic point of view; in this gelation regime, our results suggest that the observed gelation is associated with a non-equilibrium critical point, associated with a kinetic percolation phenomenon of the attaching weakly-bonded particles.

The discovery of a nonequilibrium second-order phase transition which underlies the colloidal gelation transition may open up new perspectives and opportunities for the understanding and modelling of liquid-solid transitions~\cite{Mura} in nano and soft matter systems.

\end{document}